\documentclass[10pt,letterpaper,twocolumn]{article} %% two column, final layout

\usepackage{ol2}
\usepackage[draft]{hyperref}
\usepackage{amsmath,amssymb}

\newcommand{\ve}{\varepsilon}
\newcommand{\veh}{\varepsilon_h}%{\textnormal{SiO}_2}}

\newcommand{\om}{\omega}

\newcommand{\vrt}{\scriptscriptstyle{\Vert}}

\begin{document}

\twocolumn[ %% activate for two-column option

\title{Nanoradar based on nonlinear dimer nanoantenna}

\author{Nadezhda Lapshina,$^{1,*}$ Roman Noskov,$^1$ and Yuri Kivshar$^{1,2}$}

\address{
$^1$National Research University of Information Technologies, Mechanics and Optics (ITMO), St. Petersburg 197101,  Russia\\
$^2$Nonlinear Physics Center, Research School of Physics and Engineering, \\ Australian National University, Canberra ACT 0200, Australia \\
$^*$Corresponding author: n.lapshina@phoi.ifmo.ru
}

\begin{abstract}
We introduce the concept of a nanoradar based on the operation of a nonlinear plasmonic nanoantenna. The nanoradar action originates from modulational instability occurred in a dimer nanoantenna consisting of two subwavelength nonlinear nanoparticles. Modulation instability causes a dynamical energy exchange between the nanoantenna eigenmodes resulting in periodic scanning of the nanoantenna scattering pattern. Such nanoradar demonstrates a wide scanning sector, low operation threshold, and ultrafast time
response being potentially useful for many applications in nanophotonics circuitry.

\end{abstract}

\ocis{250.5403, 230.6120, 190.3100, 320.2250.}

 ] %% activate for two-column option

The study of optical nanoantennas is a rapidly developing area of research~\cite{Biagoni}. Nanoantennas keep promises for a variety of
applications in nanophotonics with advantages of enhanced light-matter interaction. For many applications, an active control over
the nanoantenna radiation pattern is required. However, the nanoantenna operation is usually based on one lasing mode that has better directivity and/or higher Purcell factor \cite{Devilez,Krasnok}.
Nevertheless, for nanoantennas with broken material and/or geometrical symmetry, it was suggested to employ spectral tunability and variable directionality
based on switching between different modes~\cite{Bonod,Shegai} which also can be realized through nonlinear loads~\cite{Alu,Abb,Abb_2011,Maksymov}.

\begin{figure}[b]
\centerline{\includegraphics[width=8.3cm]{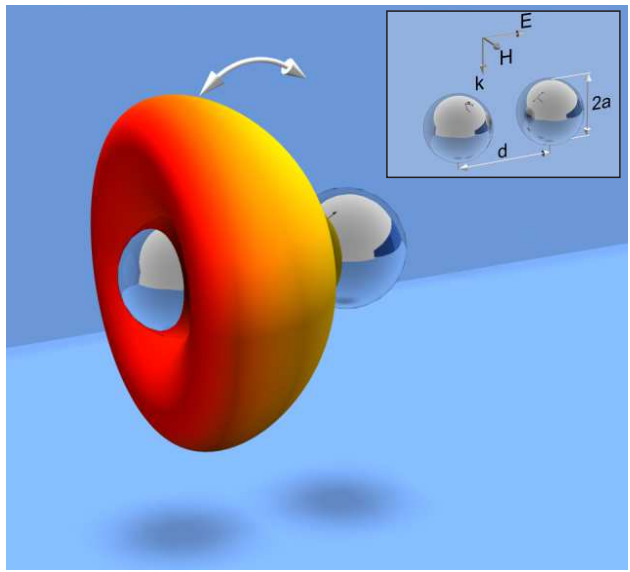}}
\caption{(Color online) Schematic of the scattering pattern of a dimer nanoantenna. Insert shows the structure, parameters, and direction of an incident plane wave.}\label{fig:1}
\end{figure}

In this Letter we suggest a novel way for the dynamic control of the nanoantenna directivity by exploiting modulational instability (MI) of a nonlinear dimer nanoantenna. We demonstrate that the development of MI in a nanodimer composed of two nonlinear silver nanoparticles can lead to a periodic exchange of the power between the eigenmodes and result in a periodic variation of the nanoantenna scattering pattern similar to classical phased-array radar systems~\cite{Jeffrey_book}, as illustrated schematically in Fig.~\ref{fig:1}.

We consider a nanoantenna created by a pair of nanoparticles placed close to each other and embedded into a SiO$_2$ host medium with a permittivity $\veh=2.15$. We assume that the nanoparticles are equivalent, and they have the radius $a=10$~nm with the center-to-center spacing of $d=30$~nm (see the insert in Fig.~\ref{fig:1}). Since the condition $a/d \leq 1/3$ is satisfied, we can employ the dipole approximation~\cite{Romero}. Assuming the nanoparticles made of silver
with a nonlinear Kerr-like response, we take the dielectric constant in the form $\ve_{\rm{Ag}}^{\rm{NL}}=\ve_{\rm{Ag}}^{\rm{L}}+\chi^{(3)}|{\bf E}^{(\rm{in})}_{1,2}|^2$, where the linear part is given by the Drude formula, $\ve_{\rm{Ag}}^{\rm{L}}=\ve_{\infty}-\om_p^2/[\om(\om-i\nu)]$, with $\ve_\infty=4.96$, $\hbar\om_p=9.54$~eV, and $\hbar\nu=0.055$~eV~\cite{Johnson}, and ${\bf E}^{(\rm{in})}_{1,2}$ are the local fields inside the first and second particles. For spherical silver nanoparticles with the radius 10 nm, we use the nonlinear coefficient with the value
$\chi^{(3)}\simeq 3\times 10^{-9}$~esu from Ref.~\cite{Drachev}, that is much larger than the cubic nonlinearity of SiO$_2$ ($\thicksim 10^{-15}$~esu~\cite{weber_book_03}), the latter is being neglected. Our aim is to study the temporal evolution of the scattering pattern of this dimer nanoantenna in the nonlinear regime when the response of this structure undergoes modulational instability.

We study a nanoparticle dimer driven by a plane wave (see the insert in Fig.~\ref{fig:1}) with the frequency close to the frequency of the surface plasmon resonance of an individual particle, and analyze the dynamical response of the particles through the evolution of their polarizations, ${\bf p}_{1,2}$. To derive the corresponding nonlinear equations for the polarizations,  we employ the model recently applied for the study of strongly localized modes in a chain of nanoparticles~\cite{noskov_prl}. In the case when the external electric field $E^{(ex)}$ is polarized along the dimer axis, this model yields the following system of coupled equations for the slowly varying amplitudes
of the particle dipole moments,
\begin{equation}
\label{dynamic}
\begin{split}
-i \frac{d P_{1}^{\vrt}}{d\tau} + \left(\Omega-i\gamma+|P_{1}^{\vrt}|^2 \right) P_{1}^{\vrt}+ G^{\vrt} P_{2}^{\vrt} = E, \\
-i \frac{d P_{2}^{\vrt}}{d\tau} +\left(\Omega-i\gamma+|P_{2}^{\vrt}|^2 \right) P_{2}^{\vrt}+ G^{\vrt} P_{1}^{\vrt} = E,
\end{split}
\end{equation}
where  the effective coefficient $G^{\vrt}  = 3\veh/(\ve_{\infty}+2\veh)\left(a/d\right)^3 \left(i k_0 d + 1 \right) \exp(-i k_0 d)$
describes the strength of the particle interaction,  and we use the dimensionless functions
 \begin{equation*}
 P_{1,2}^{\vrt}=\frac{\sqrt{\chi^{(3)}}\, p_{1,2}^{\vrt} }{\sqrt{2(\ve_\infty+2\veh)}\veh a^3},
 \end{equation*}
 and $E = -3 \veh [\chi^{(3)}]^{1/2} E^{(ex)}/[2(\ve_\infty+2\veh)]^{3/2}$ for the slowly varying amplitudes of the particle dipole moments and external electric field,
 respectively. The index '$\vrt$' stands for the longitudinal components with respect to the dimer axis,
the parameter $\gamma=\nu/(2\om_0)+(k_0 a)^3\veh/(\ve_\infty+2\veh)$ describes thermal and radiation losses of particles, $\om_0=\om_p/(\ve_\infty + 2\veh)^{1/2}$ is the frequency of the surface plasmon resonance of an individual particle, $k_0=\om_0/c\sqrt{\veh}$, $\Omega=(\om-\om_0)/\om_0$, and $\tau = \om_0 t$. For the chosen configuration $\hbar\om_0=3.14$~eV. Nonlinear terms appeared after expressing $E^{(\rm{in})}_{1,2}$ via $p_{1,2}^{\vrt}$~\cite{noskov_prl}. Equation~(\ref{dynamic}) describes the temporal nonlinear dynamics of a metallic nanodimer driven by a plane wave with the frequency $\om\sim\om_0$.

The stationary states of this system are described by homogeneous solutions when the particle dipole moments coincide, $P_{1,2}^{\vrt}=P_0^{\vrt}$; they can be found as solutions of the nonlinear equation,
\begin{equation}\label{initial state}
\left(-i\gamma+\Omega+G^{\vrt}+|P_0^{\vrt}|^2 \right) P_0^{\vrt}= E_0,
\end{equation}
where $E_0$ is the stationary amplitude of the external electric field.
This equation has multiple solutions for $\Omega<-{\rm Re}G^{\vrt} - \sqrt{3}\left( \gamma - {\rm Im}G^{\vrt} \right)$, so that the system of two nonlinear nanoparticles is expected to have a bistable regime.

\begin{figure}[t]
\centerline{\includegraphics[width=8.1cm]{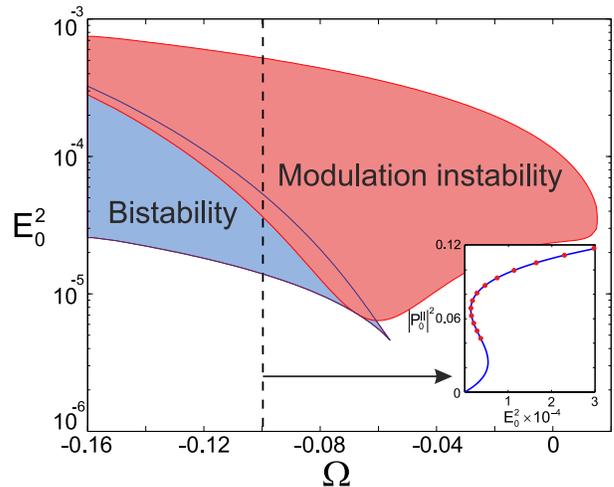}}
\caption{(Color online) Bifurcation diagram on the parameter plane of $E_0^2$ and
$\Omega=(\om-\om_0)/\om_0$, with the regions of bistability and modulation instability.
Inset: Dependence of the polarization $|P_0^{||}|^2$ on $E_0^2$ at $\Omega=-0.1$ (along the dashed vertical
line on the main plot). Red dots mark the region of modulation instability.}
\label{fig:2}
\end{figure}

Next, we analyze linear stability of the homogeneous stationary states with respect to weak perturbations taken in the form of an asymmetric longitudinal eigenmode. By applying a standard technique \cite{Rabinovich}, we
derive the expression for the instability growth rate,
\[
\lambda=-{\rm Im} G^{\vrt}-\gamma + \biggr\{|P_0^{\vrt}|^4-\biggr(2|P_0^{\vrt}|^2+\Omega-{\rm{Re}} G^{\vrt}\biggr)^2\biggr\}^{1/2}.
\]
The homogenous state (\ref{initial state}) becomes unstable provided $\lambda(\Omega,E_0^2)>0$, and the condition $\lambda=0$ defines the boundary of the MI region on the plane $(\Omega,E_0^2)$, as shown in Fig.~\ref{fig:2}. Remarkably, MI occurs for the whole upper branch in the bistable region, and it extends much further (see the inset in Fig.~\ref{fig:2}).
\begin{figure}[t]
\centerline{\includegraphics[width=8.3cm]{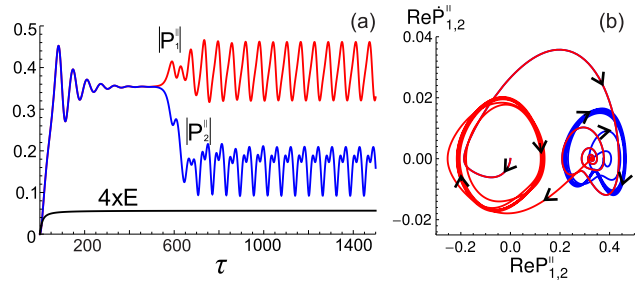}}
\caption{(Color online) (a) Temporal evolution of the polarizations $|P_1^{\vrt}|$ (red) and $|P_2^{\vrt}|$ (blue), when the external field $E$ (black)
approaches and crosses the instability threshold. Instability leads to an energy exchange between symmetric and asymmetric eigenmodes. (b) Phase trajectories associated with the dynamics shown in (a).}
\label{fig:3}
\end{figure}

To study the evolution of this nonlinear system after the onset of instability, we perform numerical simulations of Eq. (\ref{dynamic}) at zero initial conditions
and fixed driving frequency $\Omega=-0.12$ ($\hbar\om\cong2.76$~eV), when the external field
$E$ grows slowly approaching the threshold value $E_{sat}=0.014$ corresponding
to the region of MI. The results are presented in Figs.~\ref{fig:3}(a,b), where we observe that instability results in the spontaneous excitation of an asymmetric eigenmode of the dimer leading to an energy exchange between the symmetric and asymmetric eigenmodes. For weak external fields, the particle polarizations coincide being described by the phase-space trajectories corresponding to the focus. When $E$ grows reaching the region of MI, this state becomes unstable, and the phase trajectories get attracted by two different limiting circles, as shown in Fig.~\ref{fig:3}(b).

\begin{figure}[t]
\centerline{\includegraphics[width=8.3cm]{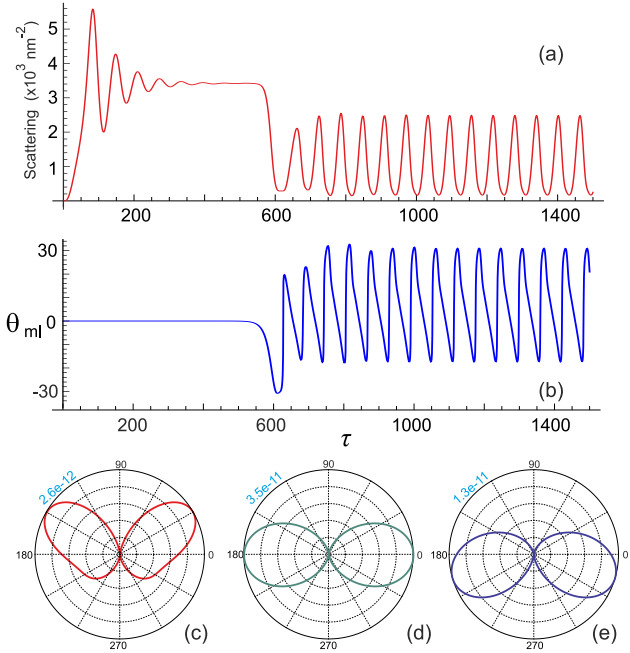}}
\caption{(Color online) Temporal evolution of (a) scattering cross-section and (b) oscillation angle $\theta_{ml}$ corresponding to the major scattering lobe. (c-e) Snapshots of the scattering intensities
(in watts) corresponding to maximal deviations from the initial state (red and blue)
and an intermediate position (green).}
\label{fig:4}
\end{figure}

MI breaks the phase locking between the particle dipole moments and external field. Consequently, self-modulation dynamical
response of a nanodimer results in temporal modulation of the nanoantenna scattering cross-section and rotation of the scattering pattern,
as shown in Figs.~\ref{fig:4}(a-e). Period of self-oscillations and contrast of scattering are $90$~fs and $18$, respectively.
This modulation is much faster than that typically observed using plasmonic antennas with nonlinear semiconductor loading~\cite{Abb_2011}.

We notice a relatively wide angular scanning sector $\theta_{ml}\simeq50^0$ of the nanodimer's radiation pattern. However, using a nanoradar in practice may require much better
directivity than the directivity of a dimer antenna, for which one may utilize, e.g., a high-permittivity dielectric sphere as a director~\cite{Devilez}.

The saturation amplitude of the external field $E_{sat}=0.014$ corresponds to the intensity of $21.6$~MW/cm$^2$. Such a high illuminating power can lead to thermal damage
 unless we operate with short pulses. To estimate the maximal pulse duration, we use the value of the ablation threshold of 3.96 J/cm$^2$ obtained for silver particles in a SiO$_2$ host matrix in the picosecond regime of illumination~\cite{Torres}. Taking into account the amplification of the electric field inside the Ag nanoparticle due to surface plasmon resonance and
 the required intensity, we evaluate the maximal pulse duration as $0.5$~ns, which is much longer than the characteristic period of self-oscillations.
 Thus, the predicted scattering pattern rotation can be readily observed in experiment.

In conclusion, we have suggested a novel approach for realizing an active control over the optical nanoantenna directivity. We have shown that modulational instability
in a nonlinear plasmonic dimer can lead to a dynamical energy exchange between the nanoantenna eigenmodes resulting in a periodical variation of
the nanoantenna scattering pattern. Such actively tunable nanoantennas may operate in a wide scanning sector with a low operation threshold and ultrafast modulation.
Further optimization may be possible by considering other designs of nanoantennas with higher directionality, such as Yagi-Uda \cite{Biagoni} or metal-dielectric \cite{Devilez} configurations.
Another possibility for tunability of the nanoradar is a variation in the plane wave incidence, intensity, and frequency. In this case, different scenarios of the development of modulational instability include  transitions to other steady states, bifurcations, period doubling, and even a transition to chaos. We have studied some of such regimes and results will be published elsewhere.

The authors thank P. Belov and A. Zharov for useful discussions, and they acknowledge a support from the Ministry of Education and Science of the Russian Federation and the Australian Research Council.

\pagebreak

\section*{Informational Fourth Page}

\bibliographystyle{osajnl}
\bibliography{References}

\end{document}